\documentclass[aps,prc,twocolumn,floatfix,showpacs,a4paper,
nofootinbib,amsmath,amssymb]{revtex4-1}
\usepackage{graphicx}
\usepackage{dcolumn}
\usepackage{bm}
\usepackage{color}

\newcommand{\be}{\begin{equation}}
\newcommand{\ee}{\end{equation}}
\newcommand{\ba}{\begin{eqnarray}}
\newcommand{\ea}{\end{eqnarray}}
\newcommand{\bd}{\begin{displaymath}}
\newcommand{\ed}{\end{displaymath}}
\newcommand{\bea}{\begin{eqnarray}}
\newcommand{\eea}{\end{eqnarray}}
\newcommand{\di}{{\rm d}}
\renewcommand{\vec}[1]{\mbox{\boldmath$#1$}}

\begin{document}
\title{$\Lambda$ polarization in peripheral collisions at moderate
relativistic energies}

\author{Y.L. Xie$^1$, M. Bleicher$^{2,3}$, H. St\"ocker$^{2,3}$, 
        D.J. Wang$^4$, and L.P. Csernai$^1$}

\affiliation{
$^1$Institute of Physics and Technology, University of Bergen,
Allegaten 55, 5007 Bergen, Norway\\
$^2$ Frankfurt Institute for Advanced Studies - Goethe University, 
60438 Frankfurt am Main, Germany\\
$^3$ Institut f\"ur Theoretische Physik, Goethe University, 
60438 Frankfurt am Main, Germany\\
$^4$ School of Science, Wuhan University of Technology, 
430070, Wuhan, China
}

\begin{abstract}
The polarization of $\Lambda$ hyperons from relativistic flow 
vorticity is studied in peripheral heavy ion reactions 
at FAIR and NICA energies, just
above the threshold of the transition to the Quark-Gluon Plasma.
Previous calculations at higher energies
with larger initial angular momentum, predicted
significant $\Lambda$ polarization based on the 
classical vorticity term in the polarization, while
relativistic modifications decreased the polarization
and changed its structure in the momentum space.
At the lower energies studied here,
we see the same effect namely that the  
relativistic modifications decrease 
the polarization arising from the initial shear flow vorticity.
\end{abstract}

\date{\today}


\maketitle

\section{Introduction}

Relativistic heavy ion collisions allow to explore the properties
of hot and dense QCD matter in the laboratory. Among the most 
prominent observables are the different kinds of transverse flow
e.g. radial flow, directed flow, elliptical flow, and higher
order flows. Hydrodynamics has been shown to provide direct access
to these flow patterns.

In recent fluid dynamical models of relativistic heavy ion reactions, 
both different fluctuating modes and global collective processes lead to
flow observables. It is important to separate or split the 
two types of flow processes from each other \cite{CseStoe2014,Floe2013}.
This separation helps to precisely analyze both processes.

In peripheral heavy ion reactions, due to the initial angular momentum,
the reaction shows
a shear flow characteristics, leading to rotation \cite{hydro1}
and even Kelvin-Helmholtz Instabilities (KHI) \cite{hydro2} in the reaction
plane, due to the low viscosity Quark-Gluon Plasma. 
This possibility was indicated
by high resolution Computational Fluid Dynamics calculations using
the PICR method. The development of these processes was studied in 3+1
dimensional (3+1D) configurations that described the energy and momentum
balance realistically
\cite{WCBS14}.
The initial state model assumed
transparency as well as stopping \cite{CK85} due to
strong attractive fields with accurate impact 
parameter and rapidity dependence in the transverse
plane \cite{M2001-2}.
It assumed an initial inter-penetration of 
Lorentz contracted slabs (in most present models considered as CGC),
and strong attractive coherent Yang-Mills fields act between
these slabs, with large string tension (according to the 
color rope model
\cite{BNK84}).

In a previous work the development of vorticity was studied under the conditions
where the viscosity is estimated to have a minimum, so
the viscous dissipation is small, \cite{Son,CKM}, and the
spherical expansion is also smaller due to the lower pressure. Thus the
initial local rotation, the vorticity drops slower.

In the PICR calculation
\cite{WCBS14},
the dynamical initial state, a
Yang-Mills field theoretical model
\cite{M2001-2} was used as in ref. \cite{CMW13},
and a longitudinal
expansion lasting $4$ fm/c from the initial impact was considered. 

The classical weighted vorticity
$\Omega_{zx}$, was calculated in the reaction [x-z] plane, 
the energy of the Au+Au  collision 
was $\sqrt{s_{NN}}=4.65+4.65$ GeV, $b=0.5 b_{max}$.
%

\begin{figure}[ht] 
\begin{center}
      \includegraphics[width=7.6cm]{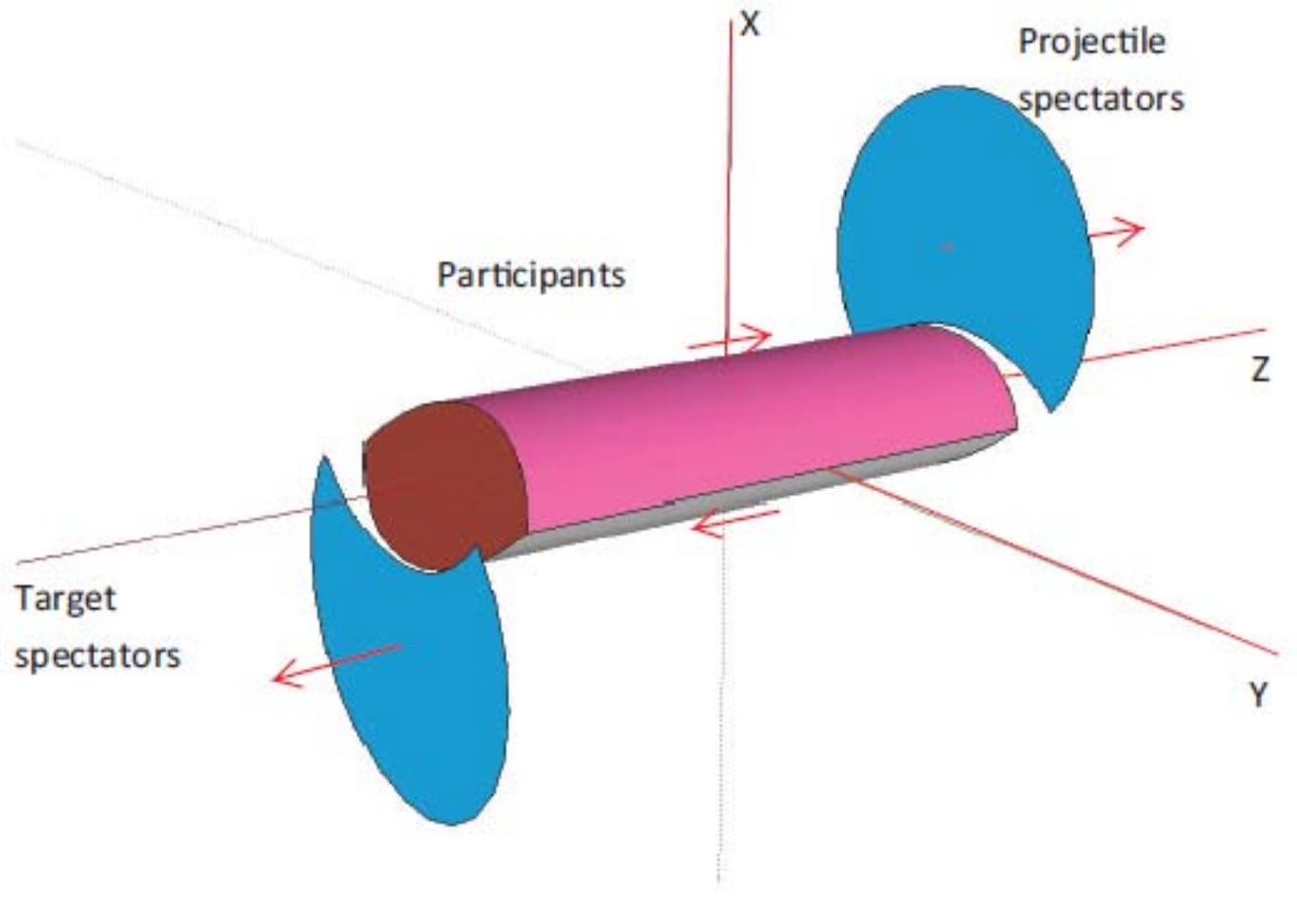}
\end{center}
\caption{
(Color online)
The three-dimensional view of the collisions
shortly after the impact. The projectile spectators are going along the z direction;
and the target spectators are going along the −z axis.The collision region is assumed
to be a cylinder with an almond-shaped
profile and tilted end surfaces, where the top side is moving to the
right and the bottom is moving to the left. The participant cylinder
can be divided into streaks, and each streak has its own velocity, as
shown in Fig. 2(a). The velocity differences among the streaks result
in rotation, turbulence and even KHI.
}
\label{P1}
\end{figure}

\begin{figure}[ht] 
\begin{center}
      \includegraphics[width=7.6cm]{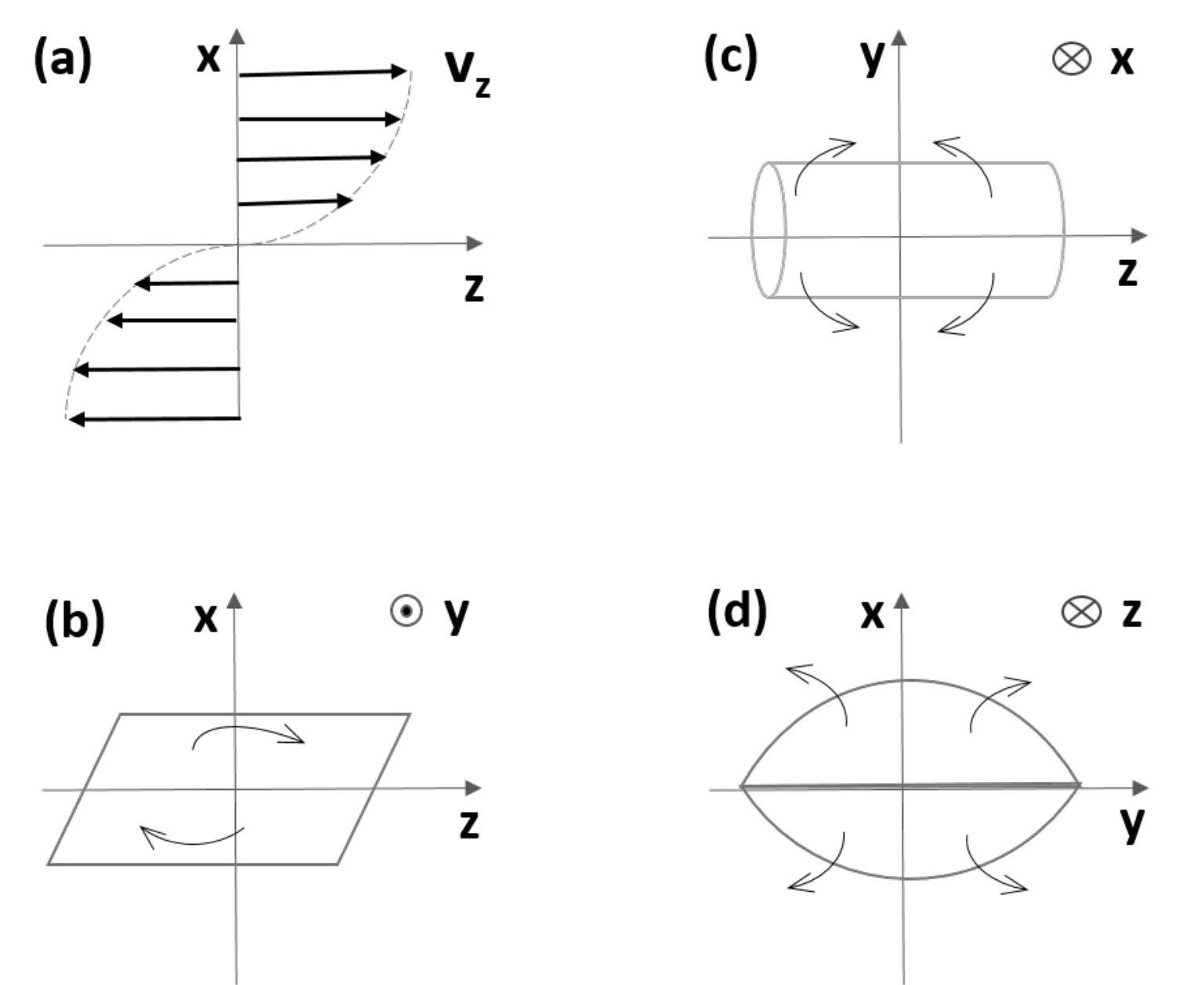}
\end{center}
\caption{
(Color online)
The schematic hydro flow velocity after the collisions shown in Fig. 1.
Panel (a) is the longitudinal velocity profile along the $x$ direction,
and it gives rise to the $v_1$ type of flow in the reaction plane, i.e. Panel (b).
Panel (c) is the anti-$v_2$ type of flow in the [y-z] plane, and Panel (d) is the 
$v_2$ type of flow in the [x-y] plane.
}
\label{P2}
\end{figure}

The used fluid
dynamical calculation and this initial state model, has been tested
in several model calculations in the last decade. It describes correctly
the initial shear flow characteristics. The angular momentum distribution is
based on the assumption that the initial angular momentum of the
participants (based on straight propagation geometry) is streak by streak
conserved, thus the model satisfies angular momentum conservation both
locally and globally.
Fig.\ref{P1} shows the 3-dimensional view of the simulated collisions shortly 
after the impact, and it could naturally generate a longitudinal velocity shear along 
the $x$ direction, as shown in Fig. \ref{P2}(a). This type of longitudinal 
velocity shear is a requirement for the subsequent rotation, turbulence and even Kelvin Helmholtz instability(KHI), just as discussed in our previous paper\cite{CKM}, as well 
as in refs. \cite{HHW11,PLB05}. The vortical flow formed in the equilibrated hydro evolution, as shown in Figs. \ref{P2}(b)(c)(d), can give rise to the polarization due to the equipartion principle or spin-orbit coupling.

The peak value of the vorticity 
at the energy $\sqrt{s_{NN}}=4.65+4.65$ GeV,
was a few times smaller
than at the ultra-relativistic RHIC and LHC energies, but the
negative values are less pronounced. The initial state used is the same
as the one that was used at high energy: we assume transparency,
QGP formation, initial longitudinal expansion in the same Yang-Mills
string rope model for $4$ fm/c time. Besides, the frequently used `Bag Model'
EoS was also applied in the hydro simulation: 
$P=c_0^2 e^2 - \frac43 B$, 
where constant $c_0^2 = \frac13$ and $B$ is the Bag constant in QCD
 \cite{M2001-2, LPS94}.
The energy density takes the form:
$e=\alpha T^4+ \beta T^2 + \gamma + B $, where $\alpha$, $\beta$, $\gamma$ 
are constants aring from the degeneracy factors for (anti-)quarks and gluons. 
At later time, the drop of the vorticity is not as large as in higher energy 
heavy ion collisions.

In ref. \cite{WCBS14} the classical and relativistic weighted vorticities,
$\Omega_{zx}$,
were evaluated in the reaction plane, [x-z],
so that the weighting does not change the average circulation
of the layer, i.e.,
the sum of the average of the weights over all fluid cells is unity.
The vorticity         projected to
the reaction plane for a collision for the FAIR-SIS300 energy of
$\sqrt{s_{NN}}=8.0$ GeV is evaluated at an initial moment of time and
at a later time.
The peak value of the vorticity is similar to the one obtained
at the ultra-relativistic RHIC and LHC energies, but the
negative values are less pronounced. 
The average vorticity was decreasing with time:
$\Omega_{zx}$  is 0.1297 / 0.0736 for the times,
$t=$0.17 and 3.56 fm/c respectively. The same behavior was 
seen in ref. \cite{JLL16}.

In addition to the directed flow ($v_1$) \cite{hydro1,Ste2014}
two methods were proposed so far to detect the effects of rotation: 
the Differential HBT method \cite{DHBT} and the polarization of
emitted fermions based on the equipartition of the rotation
between the spin and orbital degrees of freedom \cite{Bec13,BCDG13}.

The particle polarization effect has some advantages and disadvantages.
The local polarization depends on the thermal vorticity
\cite{Bec13,BCDG13}.
Now at lower collision energy the temperature is lower and the
thermal vorticity increases, which is advantageous. At ultra-relativistic
energies this feature led to the conclusion that the predicted polarization
is bigger for RHIC than for LHC, because of the lower temperature of the
system. Furthermore at ultra-relativistic energies, the relativistic 
corrections to polarization will become stronger compared to the
original shear and the resulting classical vorticity \cite{Erratum}.

\begin{figure}[ht] 
\begin{center}
      \includegraphics[width=7.6cm]{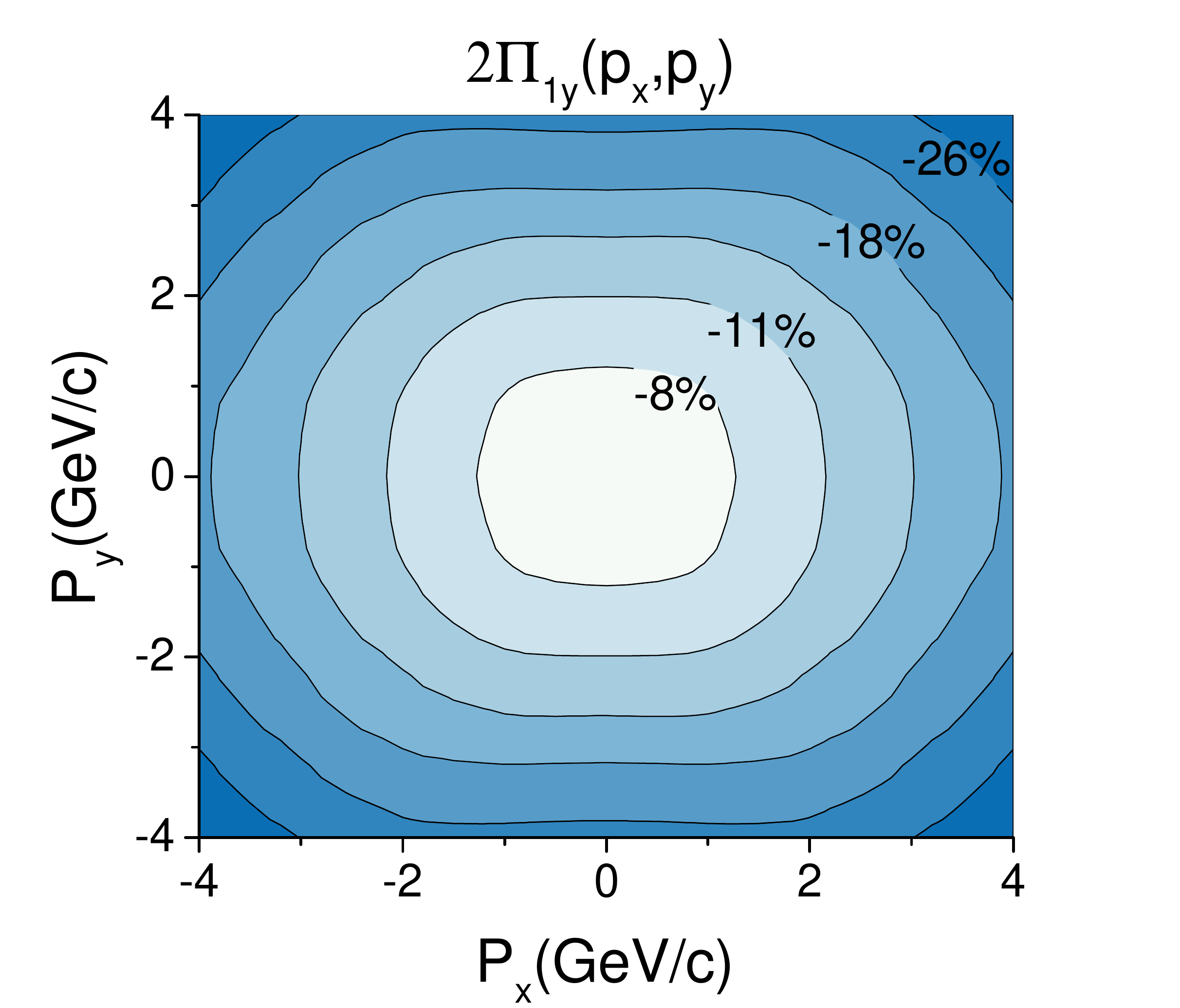}
      \includegraphics[width=7.6cm]{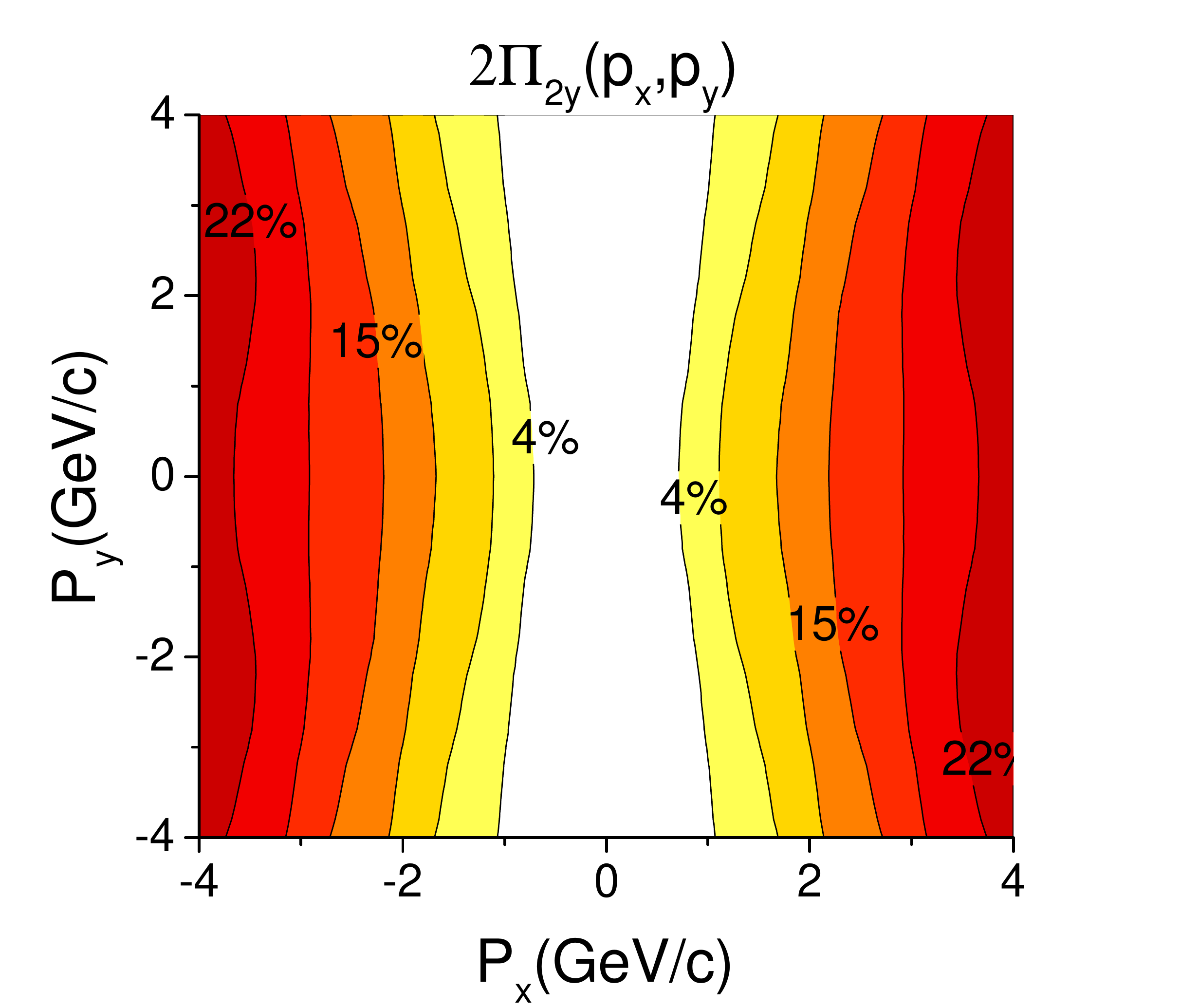}
\end{center}
\caption{
(Color online)
The first (top) and second (bottom) term of the dominant $y$ component
of the $\Lambda$ polarization
for momentum vectors in the transverse, $[p_x,p_y]$, plane at $p_z=0$,
for the FAIR U+U reaction at $\sqrt{s_{NN}}=8.0$ GeV.
}
\label{F1ab-12y}
\end{figure}

The thermal vorticity occurs in the particle polarization, because
the spin-orbit interaction aligns the spins and the orbital momentum,
while the random thermal motion works against this alignment.
Thus, we use
the inverse temperature four-vector
field \cite{Bec13,BCDG13},
$$
\beta^\mu(x) = (1/T(x)) u^\mu(x) \ ,
$$
and define the {\em thermal vorticity} as:
\be\label{thervor2}
  \varpi^{\mu \nu}=\frac{1}{2} (\partial^{\nu}\hat{\beta}^{\mu}-
  \partial^{\mu}\hat{\beta}^{\nu}),
\ee
where $\hat{\beta}^{\mu} \equiv \hbar \, \beta^{\mu}$. Thereby, 
$\varpi$ becomes dimensionless.

The relativistic weighted thermal vorticity
$\Omega_{zx}$, calculated in the reaction [x-z] plane was 
presented in ref. \cite{WCBS14}.
The  energy of the Au+Au collision was $\sqrt{s_{NN}}=4.65+4.65$ GeV, and the
impact parameter $b=0.5 b_{max}$.
The obtained average thermal vorticity,
$\Omega_{zx}$,  was 0.0847 (0.0739) for the times,
$t=$0.17 and 3.56 fm/c respectively.
It was observed that the thermal vorticity decreases 
slower than the standard vorticity
due to the decreasing temperature.

\begin{figure}[ht] 
\begin{center}
      \includegraphics[width=7.6cm]{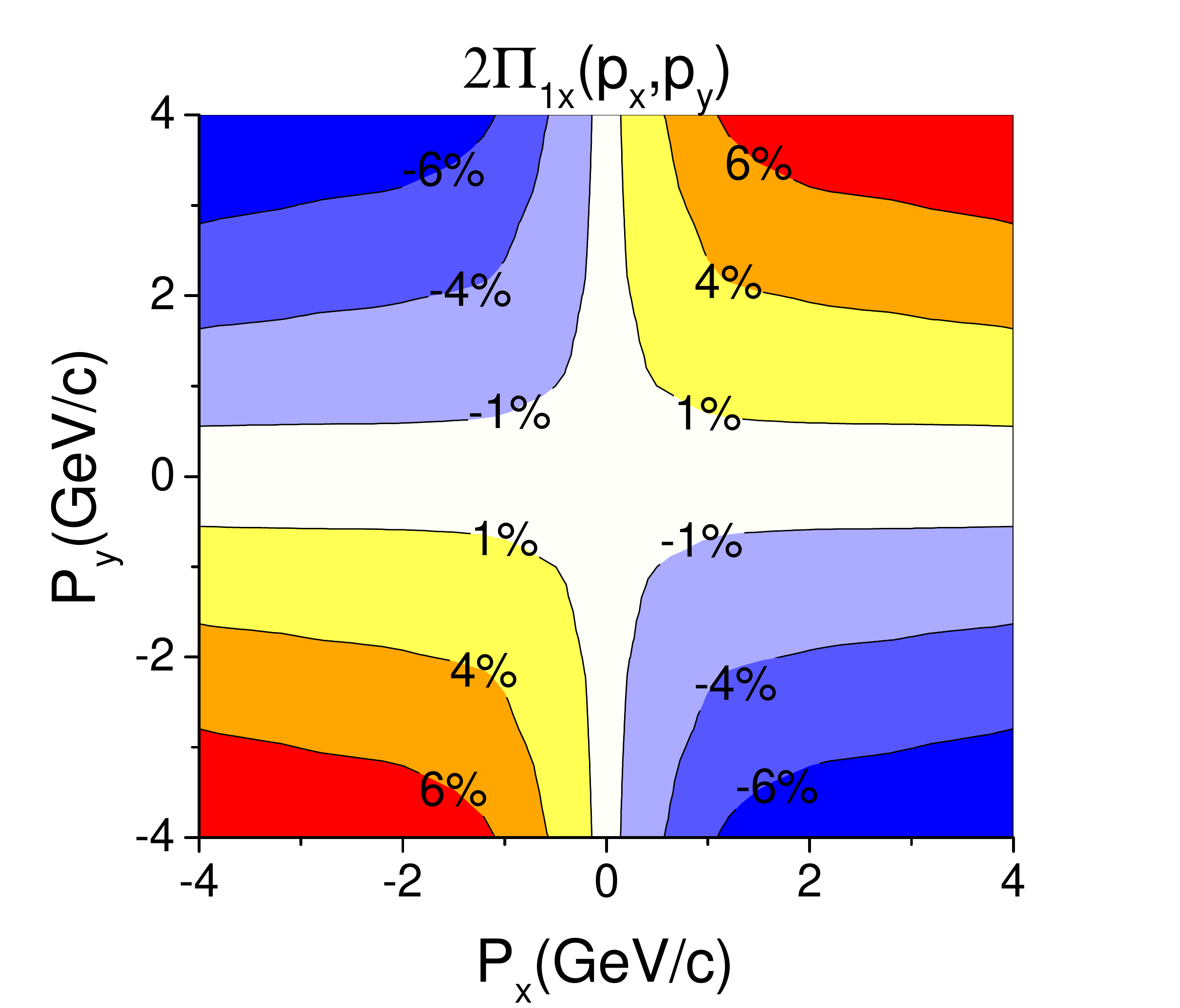}
      \includegraphics[width=7.6cm]{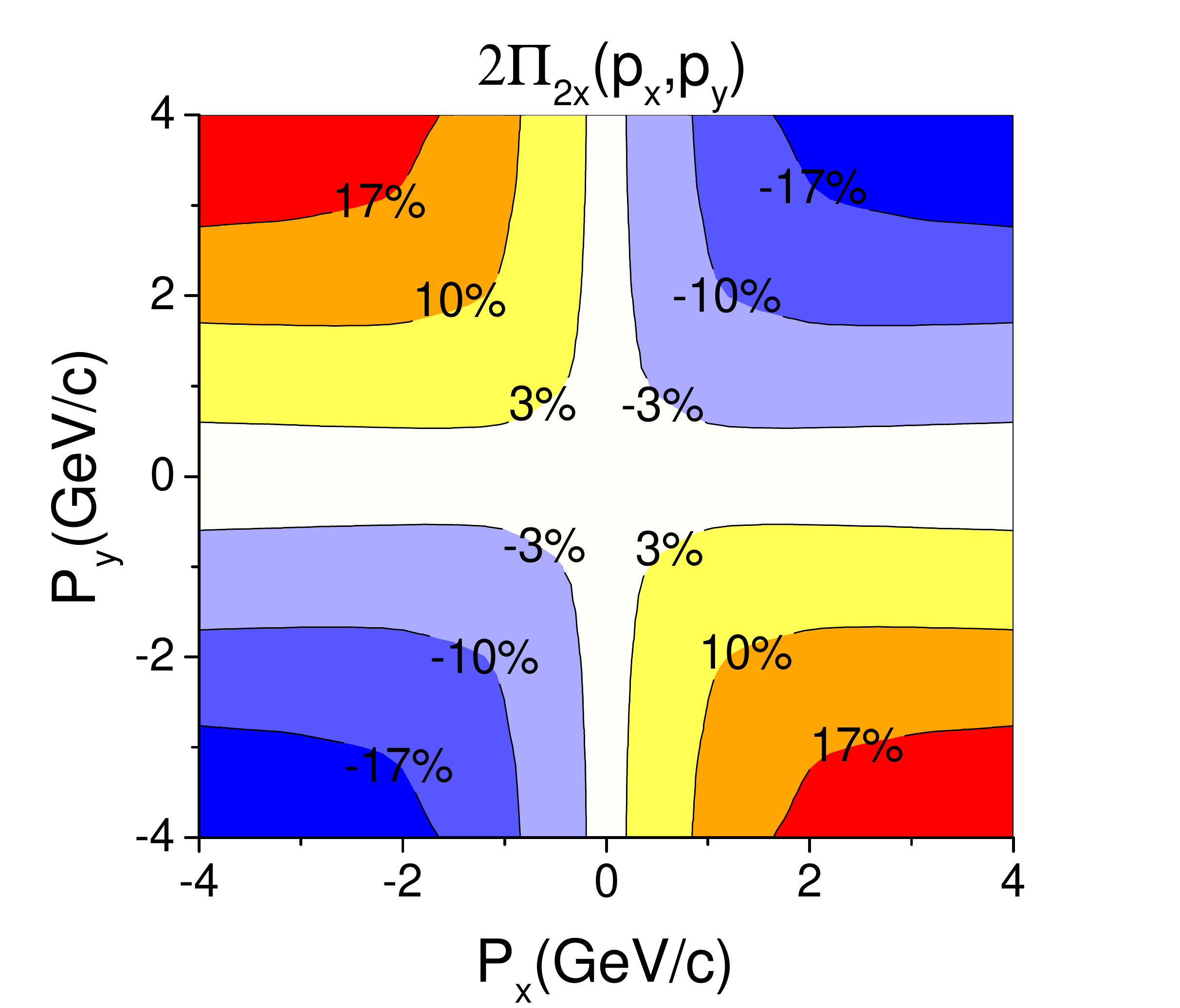}
\end{center}
\caption{
(Color online)
The first (top) and second (bottom) term of the $x$ component
of the $\Lambda$ polarization
for momentum vectors in the transverse, $[p_x,p_y]$, plane at $p_z=0$, 
for the FAIR U+U reaction at $\sqrt{s_{NN}}=8.0$ GeV.
}
\label{F2ab-12x}
\end{figure}

In ref. \cite{WCBS14}
the relativistic weighted thermal vorticity
$\Omega_{zx}$, was   calculated in the reaction [x-z] plane at t=0.34 fm/c
and at t=3.72 fm/c for
the  energy of the collision $\sqrt{s_{NN}}=4.0+4.0$ GeV,
$b=0.5 b_{max}$.
$\Omega_{zx}$  was 0.0856 (0.0658) for the two selected times.

An analysis of the vorticity for peripheral
Au+Au reactions at NICA and U+U reactions at FAIR energies of
$\sqrt{s_{NN}}=9.3 (8.0)$ GeV respectively gave
an initial peak vorticity that was about two times larger
than the one obtained from random fluctuations in the transverse
plane, of about 0.2 c/fm at much higher energies
\cite{Stefan}. This is due to the initial angular
momentum arising from the beam energy in non-central collisions.

The RHIC Beam Energy Scan program measured significant
$\Lambda$ and $\bar{\Lambda}$ polarizations, with the largest 
values at the lowest energies \cite{Lisa2016}.

\begin{figure}[ht] 
\begin{center}
      \includegraphics[width=7.6cm]{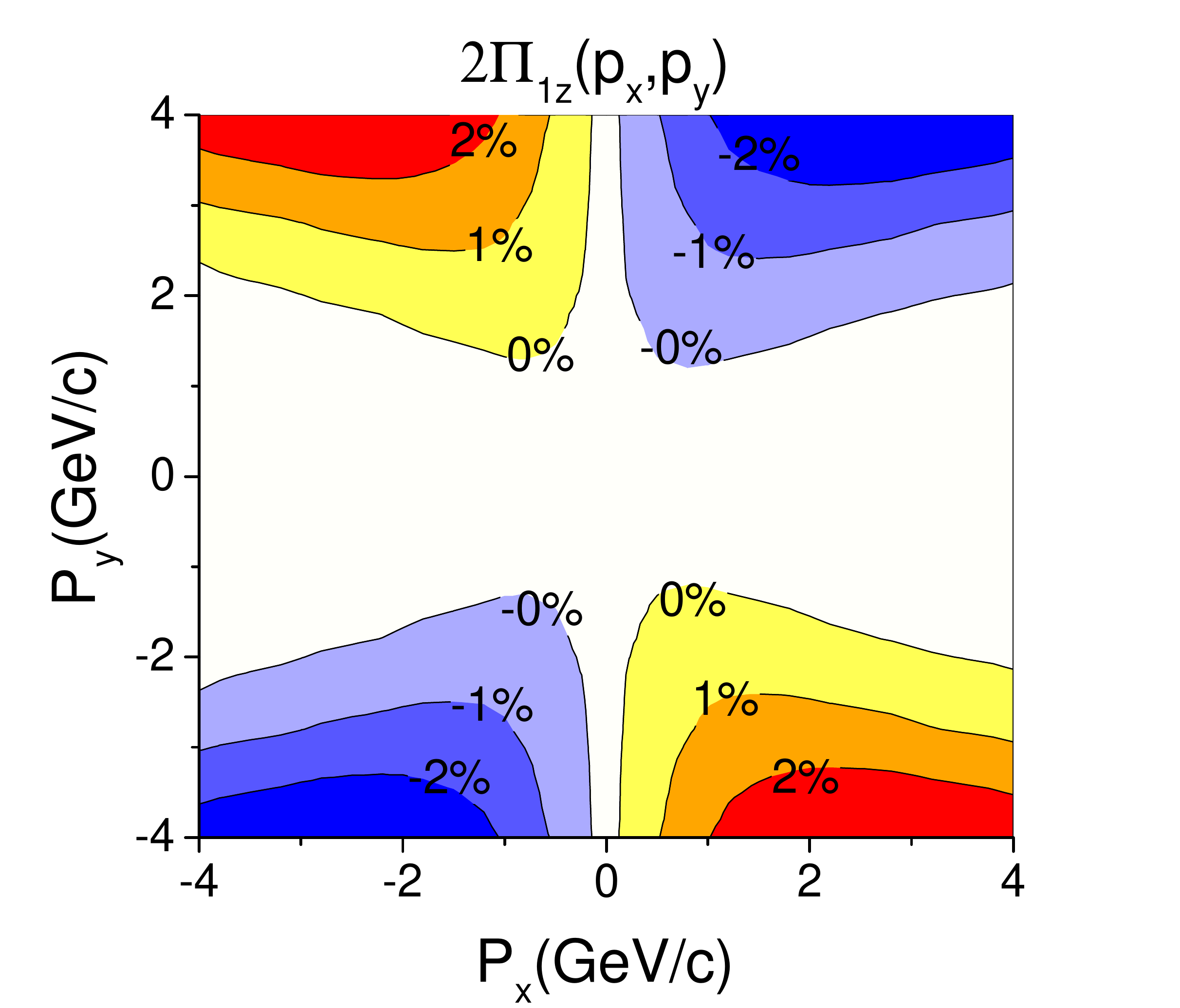}
      \includegraphics[width=7.6cm]{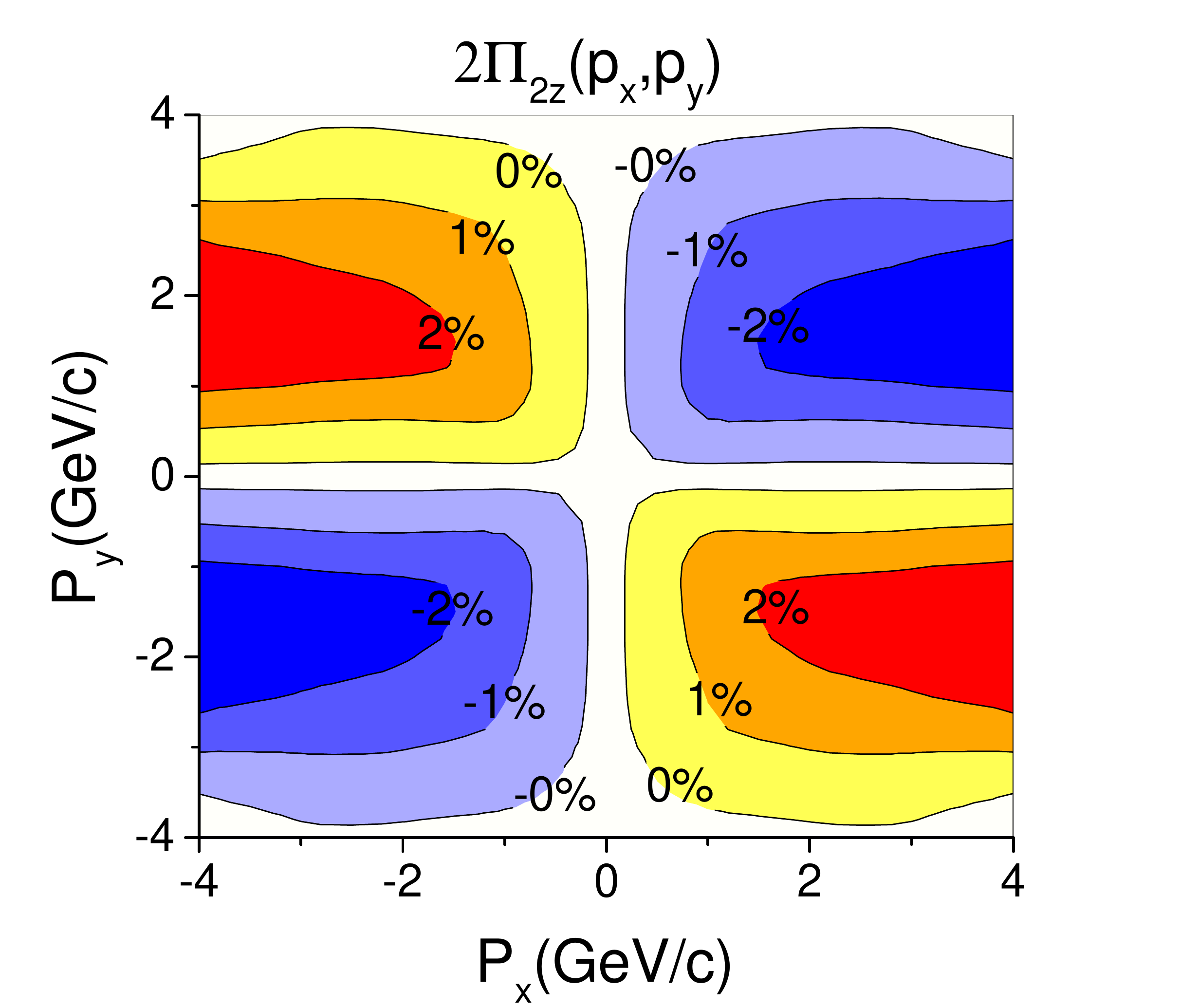}
\end{center}
\caption{
(Color online)
The first (top) and second (bottom) term of the $z$ component
of the $\Lambda$ polarization
for momentum vectors in the transverse, $[p_x,p_y]$, plane at $p_z=0$, 
for the FAIR U+U reaction at $\sqrt{s_{NN}}=8.0$ GeV.
}
\label{F3ab-12z}
\end{figure}

At FAIR, the planned facilities, e.g. at PANDA \cite{Db13},
will make it possible to measure proton and anti-proton polarization,
also
in the emission directions where significant polarization is
expected.

\section{Polarization studies}

The flow vorticity was evaluated and reported in 
\cite{WCBS14}. Based on these results we report the $\Lambda$ polarization
results for the same reactions. The initial state 
Yang-Mills flux-tube model \cite{M2001-2} describes the development
from the initial touching moment up to 2.5 fm/c. Then the PICR Hydro 
code is calculated for another 
4.75 fm/c, so that the final freeze out time is 7.25 fm/c.

\begin{figure}[ht] 
\begin{center}
      \includegraphics[width=7.6cm]{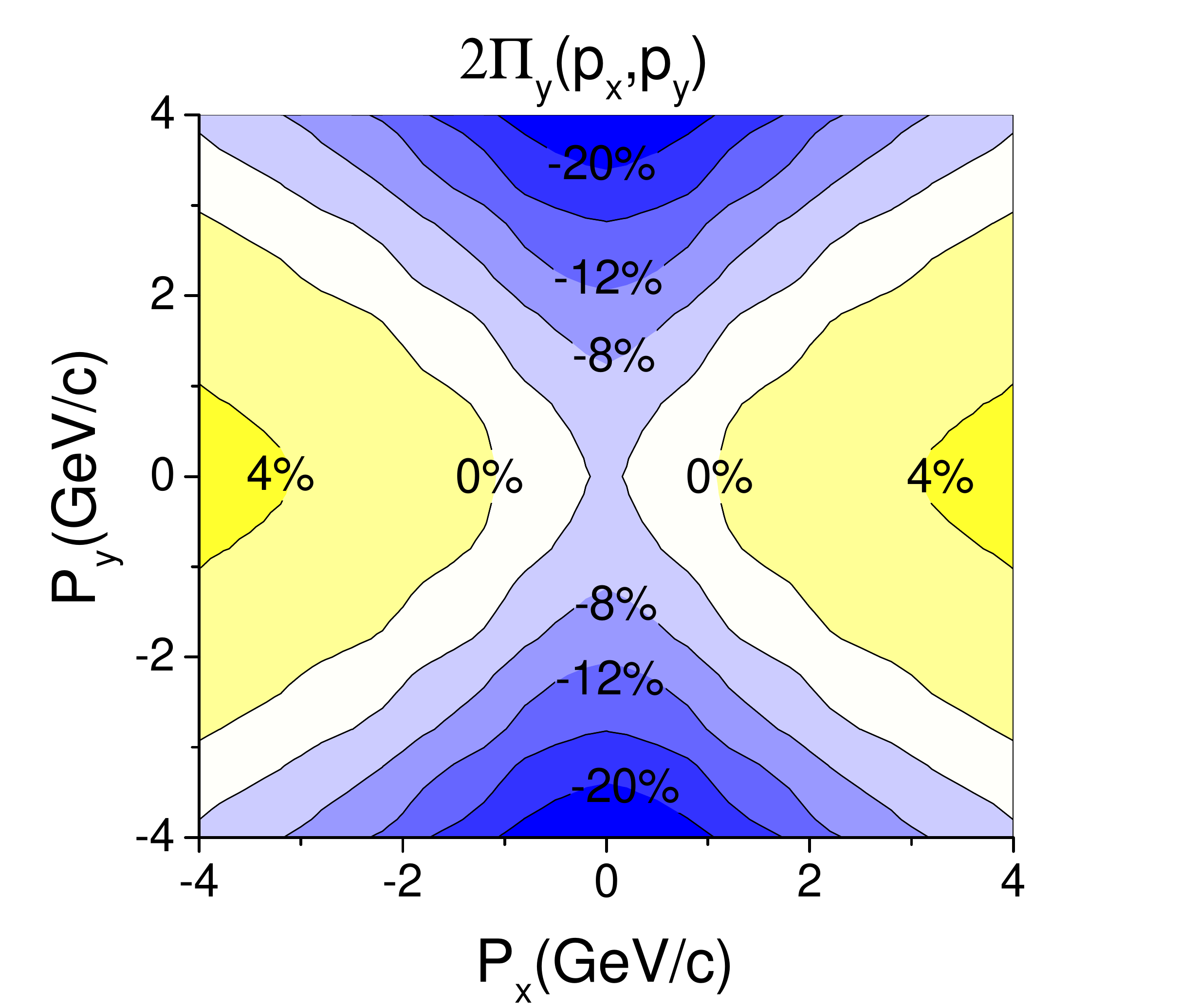}
      \includegraphics[width=7.6cm]{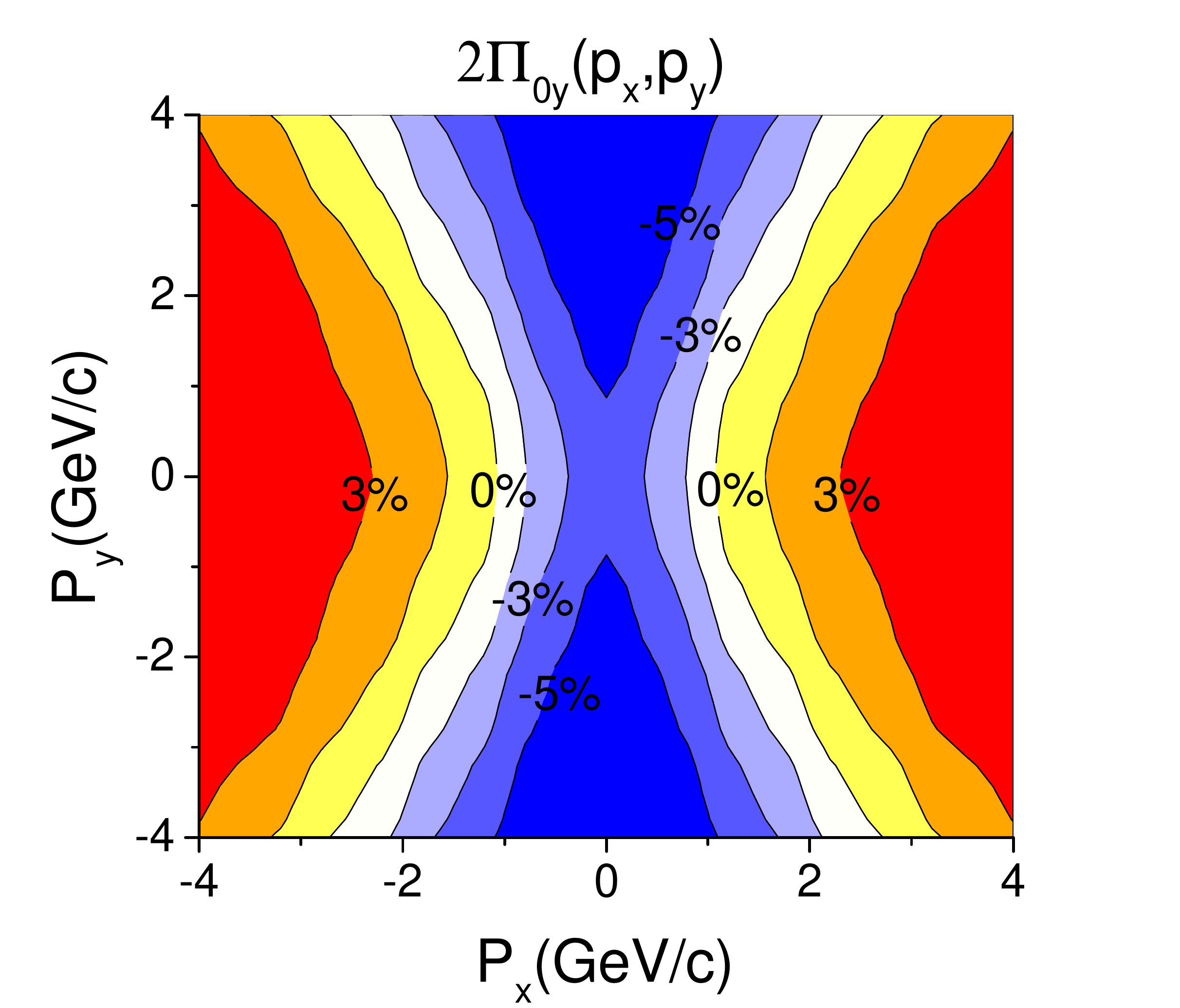}
\end{center}
\caption{
(Color online)
The $y$ component of the $\Lambda$ polarization
for momentum vectors in the transverse, $[p_x,p_y]$, plane at $p_z=0$, 
for the FAIR U+U reaction at $\sqrt{s_{NN}}=8.0$ GeV.
The top figure is in the calculation frame,  while the
bottom figure is boosted to the frame of the $\Lambda$ 
\cite{Bec13}.
}
\label{F4ab-y0y}
\end{figure}

The  $\Lambda$ and  $\bar{\Lambda}$ polarization was calculated based on
the work \cite{BCDG13}
\bea\label{Pipv}
 && \vec{\Pi}(p) = \frac{\hbar \varepsilon}{8m}
  \frac{\int \di \Sigma_\lambda p^\lambda \, n_F\ 
  (\nabla\times\vec{\beta})}{\int \di \Sigma_\lambda p^\lambda \,n_F} 
  \nonumber \\
 && + \frac{\hbar {\bf p}}{8m} \times 
  \frac{\int \di \Sigma_\lambda p^\lambda \,
 n_F\ (\partial_t \vec{\beta} + \nabla\beta^0)}
 {\int \di \Sigma_\lambda p^\lambda \,n_F} \,.
\eea
where, $n_F(x,p)$ is the Fermi-J\"uttner distribution of the $\Lambda$, 
that is $1/(e^{\beta(x)\cdot{p}-\xi(x)}+1)$, being 
$\xi(x)=\mu(x)/T(x)$ with $\mu$ the
relevant $\Lambda$ chemical potential and $p$ its four-momentum.  
$\di \Sigma_\lambda$ is the freeze out hypersurface element, for
$t=$const. freeze-out
$\di \Sigma_\lambda p^\lambda\ \to \di V \varepsilon$,
where $\varepsilon = p^0$ being the $\Lambda$'s energy.

Here the {\it first term} is the classical vorticity term, while the 
{\it second term} is the relativistic modification. 
The above convention of $\vec{\Pi}(p)$ \cite{BCDG13} is normalized
to max. 50\%, while in the experimental evaluation it is
100 \%, thus we present the values of $2 \vec{\Pi}(p)$,  
\cite{Erratum}
unlike in earlier calculations \cite{Bec13,XGCs2015,BIR15}.

In Fig. \ref{F1ab-12y} 
the dominant $y$ component of the polarization vector $\vec{\Pi}(p)$, 
for the first and second terms are shown. 
The first term is pointing into the negative $y$ direction with
a maximum of -26\%. The structure of first term arises from
the $v_1$ type of flow in Fig. \ref{P2}(b), which is also unipolar and negative y
directed. The second term has different structure, it points 
in the opposite direction and has a maximum of +22\%, i.e./ $\sim 4$\%
less than the absolute value of the first term. 

In Fig. \ref{F2ab-12x}
the $x$ component of the polarization vector $\vec{\Pi}(p)$, for
the first and second terms are shown. 
The first term is about four times smaller than the $y$ component, 
$\pm 6$\%, and the positive and negative values are symmetric in a way 
that the integrated value of the polarization over the momentum space in 
the transverse plane is vanishing.
This sign distribution is just the manifestation of anti-$v_2$ type of flow in 
[y-z] plane, seen in Fig. \ref{P2}(c) with a dipole structure.
The second term is about half of the $y$ component, 
$\pm 17$\%, and the positive and negative values are symmetric in a way 
that the integrated value of the polarization over the momentum space in 
the transverse plane is vanishing. Furthermore the first and second terms
have opposite signs at the same momentum values in the transverse plane,
which decreases further their effect.

\begin{figure}[ht] 
\begin{center}
      \includegraphics[width=7.6cm]{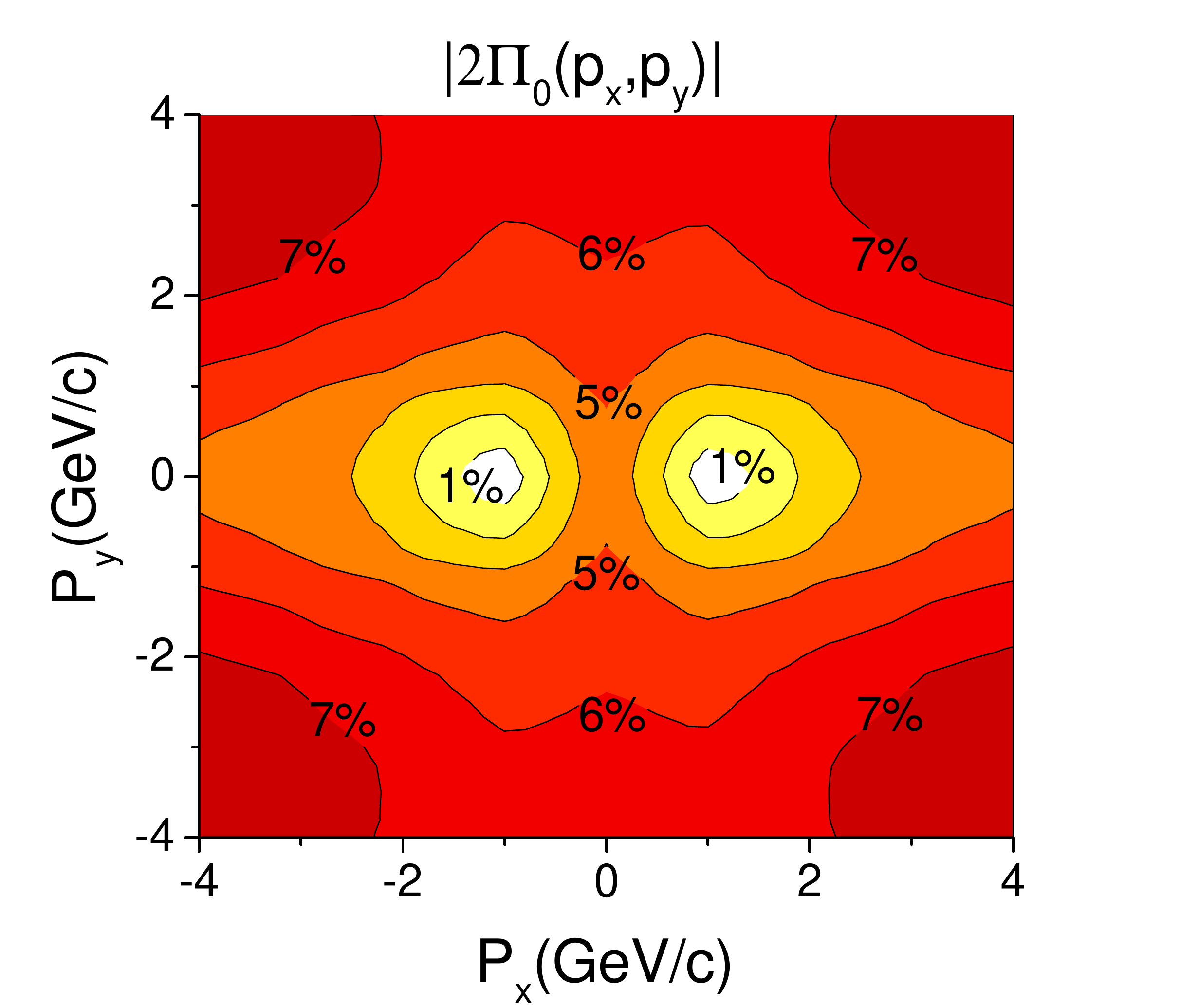}
\end{center}
\caption{
(Color online)
The modulus of the $\Lambda$ polarization
for momentum vectors in the transverse, $[p_x,p_y]$, plane at $p_z=0$, 
for the FAIR U+U reaction at $\sqrt{s_{NN}}=8.0$ GeV.
The figure is in the frame of the $\Lambda$.
}
\label{F5-0}
\end{figure}

In Fig. \ref{F3ab-12z} 
the $z$ component of the polarization vector $\vec{\Pi}(p)$, 
for the first and second terms are shown. 
The first term has a maximum of $\pm 2$\%,
and the positive and negative values are symmetric in a way 
that the integrated value of the polarization over the momentum space in 
the transverse plane is vanishing.
This sign distribution is also the manifestation of the anti-$v_2$ type of 
flow in [x-y] plane, i.e. a dipole structure in Fig. \ref{P2}(d).
The second term has similar structure to the first one, with a 
maximum of $\pm 2$\% also, but the first and second terms have similar
structure in the momentum space.

In Fig. \ref{F4ab-y0y} 
the dominant $y$ component of the polarization vector $\vec{\Pi}(p)$, 
for the sum of the first and second terms is shown. 
The top figure is the distribution of the polarization in the
center-of-mass frame while the bottom figure is in the local rest frame
of the $\Lambda$.

\begin{figure}[ht] 
\begin{center}
      \includegraphics[width=7.6cm]{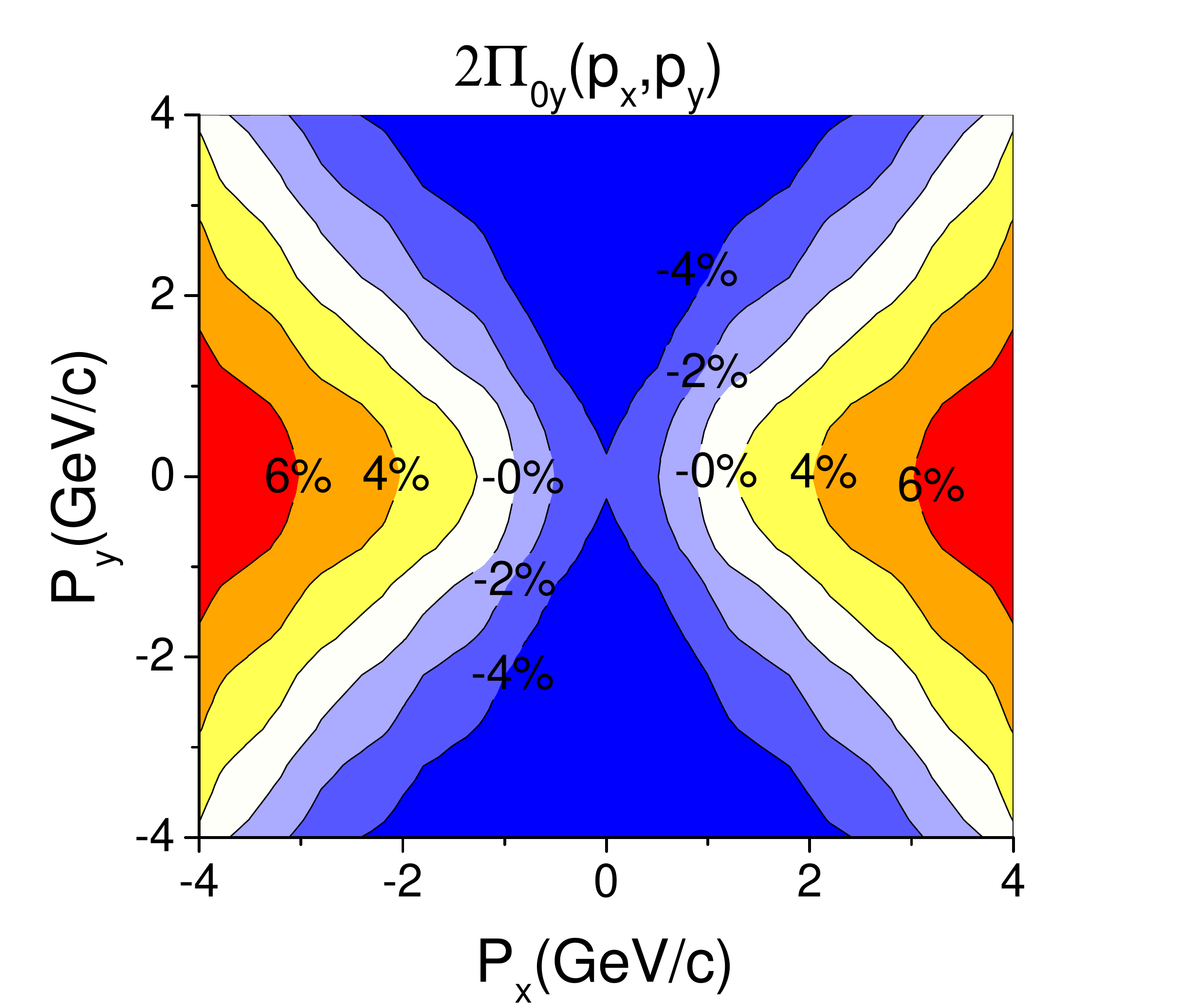}
      \includegraphics[width=7.6cm]{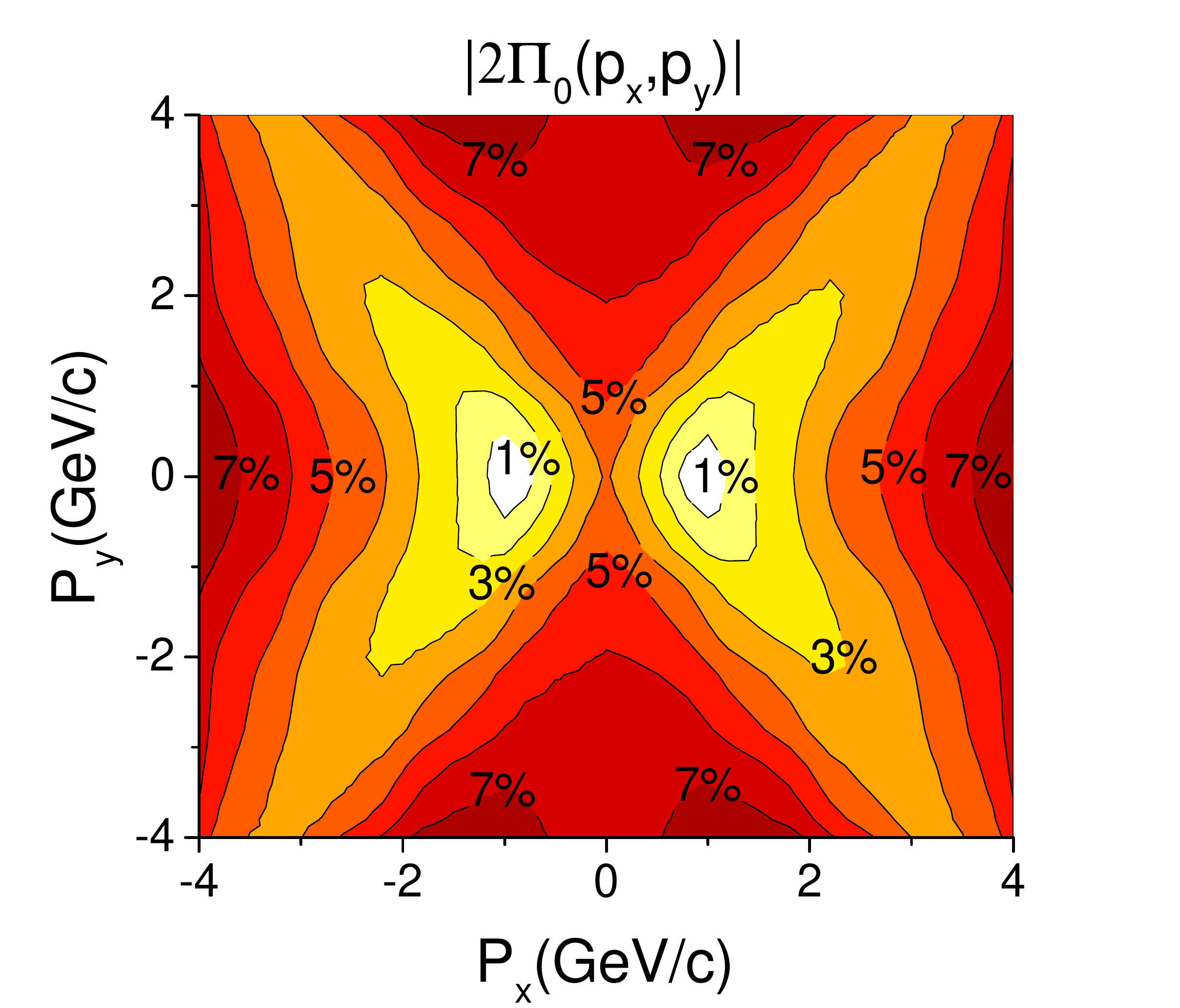}
\end{center}
\caption{
(Color online)
The $y$ component (top) and the modulus (bottom) of the $\Lambda$ 
polarization for momentum vectors in the transverse, 
$[p_x,p_y]$, plane at $p_z=0$, 
for the NICA Au+Au reaction at $\sqrt{s_{NN}}=9.3$ GeV.
The figure is in the frame of the $\Lambda$.
}
\label{F6ab-0y0}
\end{figure}

Fig. \ref{F5-0}
shows the modulus of the polarization vector $\vec{\Pi}(p)$.
The maximum at high $|p_y|$ and low $|p_x|$ is the same as the
absolute value of the  ${\Pi}_{0y}$
component. Here the other components have only minor 
contributions to the final observed polarization. At the
corners, at high $|p_y|$ and high $|p_x|$, the contribution of
the $x$ and $z$ components of $\vec{\Pi}(p)$ dominates,
while the $y$ component has a minimum.

Fig. \ref{F6ab-0y0} shows the $y$ component and 
the modulus of the polarization vector $\vec{\Pi}(p)$
for the NICA Au+Au reaction at $\sqrt{s_{NN}}=9.3$ GeV.
The structure and magnitude of the polarization is
similar to the reactions at FAIR.  
The negative maximum at high $|p_y|$ and low $|p_x|$ arises
from the classical vorticity in the $y$ component. 
The positive maximum at high $|p_x|$ and low $|p_y|$ arises
from the relativistic modifications of the second term.
The momentum space average is dominated by the first term.

\begin{figure}[h] 
\begin{center}
      \includegraphics[width=7.6cm]{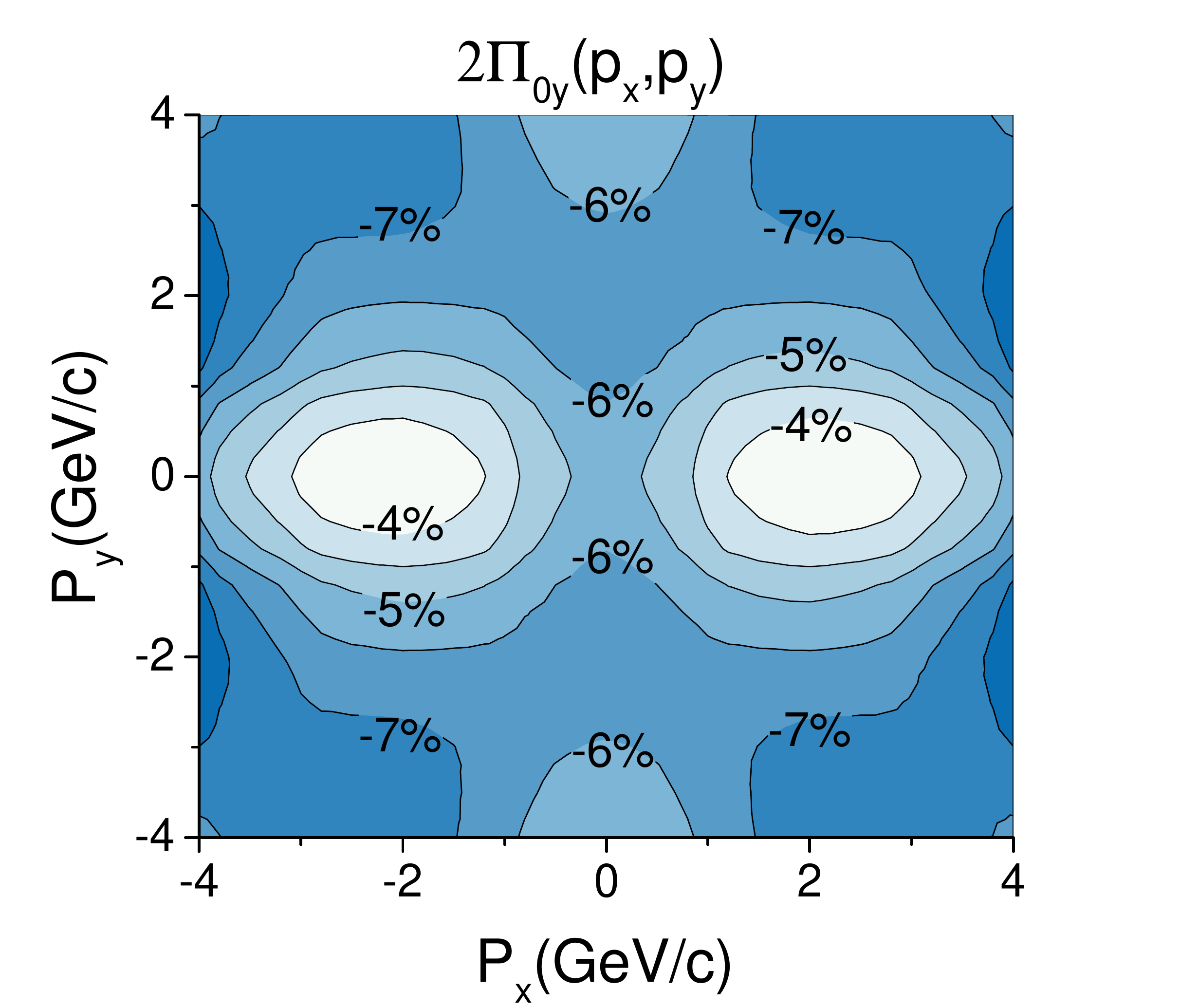}
      \includegraphics[width=7.6cm]{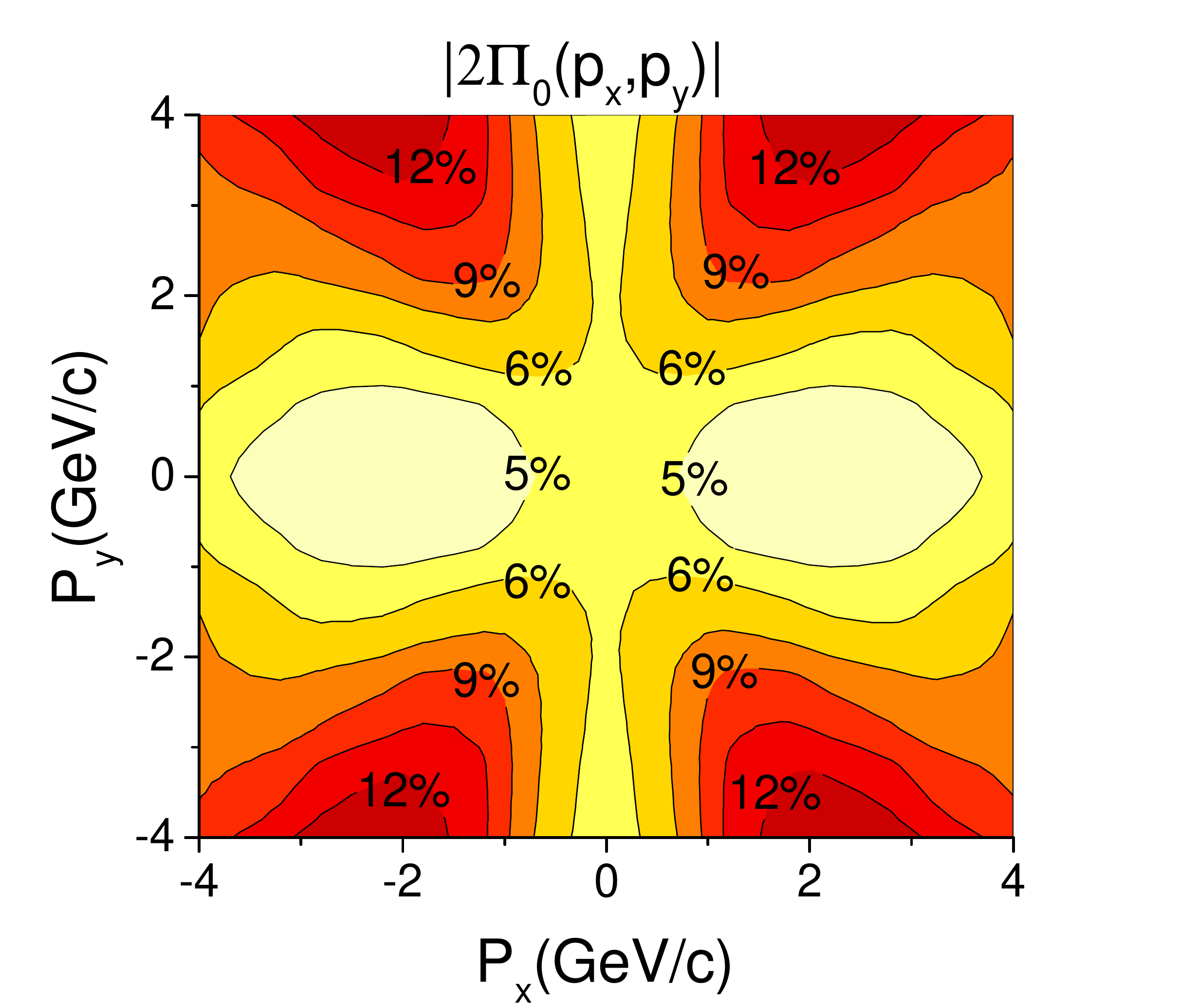}
\end{center}
\caption{
(Color online)
The $y$ component (top) and the modulus (bottom) of the $\Lambda$ 
polarization for momentum vectors in the transverse, 
$[p_x,p_y]$, plane at $p_z=0$, 
for the FAIR U+U reaction at $\sqrt{s_{NN}}=8$ GeV at the
earlier freeze out time of $t=4.2$ fm/c.
The figure is in the frame of the $\Lambda$.
}
\label{F7ab-0y0}
\end{figure}

The polarization studies at ultra-relativistic, RHIC and LHC energies,
turned out to be sensitive to both the classical vorticity of the
flow (first term) and the relativistic modifications arising from rapid
expansion expansion at later stages of the flow (second term)
\cite{Bec13,Erratum}.

Initially the contribution of the classical vorticity is stronger than
the relativistic modification term, i.e. the `second'  term. This is in line with
earlier observations \cite{WCBS14,JLL16}. The effect of this decrease is also
visible in the polarization results. The $\Lambda$ polarization
was evaluated at earlier freeze out time, $t= 2.5 + 1.7$fm/c $= 4.2$fm/c
for the FAIR U+U reaction.
See Fig. \ref{F7ab-0y0}

The $y$ component and the modulus of the polarization vector $\vec{\Pi}(p)$
have very similar structure and magnitude, although the $y$ component
points in the negative $y$ direction as the angular momentum vector from the 
initial shear flow. This indicates that the other, $x$ and $z$, components
are of the order of 1\% only at moderate momenta where the $y$ component
and the modulus are of the order of 5-6\%. At the "corners",
at high $|p_y|$ and high $|p_x|$, the contribution of
the $x$ and $z$ components of $\vec{\Pi}(p)$ are approaching that of the
$y$ component, so that the modulus is larger than the $y$ component, by
4-5\%. Still the contribution of these second term components is 
clearly smaller than the classical vorticity component. 

It is important to mention the role of the initial condition. 
The {\it second term}, the relativistic modification, develops
during the expansion of the system and is not very sensitive to the 
initial state. This is shown by the fact that the structure of the 
$x$ component of polarization, $\Pi_{2x}$ in the dominant 
Fig. \ref{F2ab-12x}b, is very 
similar to Fig. 14b of ref. \cite{BIR15}. 
At the same time here the
initial shear and classical vorticity are present in the initial state
with strong stopping and dominance of the Yang-Mills field, 
\cite{CK85,M2001-2}, while in ref. \cite{BIR15} this is not present.
As a consequence the final polarization estimates in the $y$ direction
are different in the two models.

\section{Total $\Lambda$ Polarization integrated over momentum space}

Since the experimental results for $\Lambda$ polarization are
averaged polarizations over the $\Lambda$ momentum, we evaluated the
average of the $y$ component of the polarization 
$\langle\Pi_{0y}\rangle_{p}$.
We integrated the  $y$ component of the obtained
polarization, $\Pi_{0y}$, over the momentum space as follows:
\ba
\langle\Pi_{0y}\rangle_{p}&=&
\frac{\int dp\, dx \,\Pi_{0y}(p,x) \, n_F(x,p)}{\int dp\, dx\, n_F(x,p)} 
\nonumber \\
&=&\frac{\int \,dp \,\Pi_{0y}(p)\, n_F(p) }{ \int \,dp \,n_F(p)}
\ea

For Au+Au collisions at NICA energy (9.3 GeV/A), the avarged $y$ 
component of polarization is 1.82\%,
at freeze out time 2.5+4.75 fm/c,  
while for the U+U collisions at FAIR energy (8 GeV/A) at the same time,
the value is 1.85\%, 
a bit larger. 
As some papers 
\cite{Bec13,AM87}
had pointed out, the $\Lambda$ polarization scales with
 $x_F \, = \, 2p/\sqrt{s}$, 
thus the $\Lambda$ polarization should increase with
decreasing energy, which is also, more or less, being confirmed 
by our results. We also evaluated the
average polarization for U+U collisions at 8 GeV/A energy at
an earlier time 2.5+1.7 fm/c, and 
the obtained value is about 2.02\%, 
showing that the average polarization is decreasing
with freeze out time.

It is important to mention that if vorticity and polarization
are dominated by the expansion and not by the initial shear flow then
the vorticity is symmetric, see Fig. 13 of ref. \cite{BIR15}, and the
polarization also as shown in Fig. \ref{F2ab-12x}b, and similarly 
in Fig. 14b of ref. \cite{BIR15}. Due to the symmetry of the polarization
in the $\pm$ directions these polarizations average out to zero.
This applies to the $y$ directed polarization in the model initial 
state of ref. \cite{BIR15} also. The result that the present model 
yields to a net average polarization, ${\Pi}_{0y})$, 
in the negative $y$ direction, is due to the strong shear 
flow and vorticity in the initial state.

The vorticity induced by the initial orbital angular momentum will 
eventually give rise to non-vanishing local and global polarization, 
which is aligned with the initial angular momentum 
\cite{GLP12,HHW11,GCD08,PLB05,PRL05,AP08,PangEA2016}. 
As eq. (4.4) in Ref. \cite{HHW11} shows, the quark polarization 
rate is sensitive to the viscosity, $\eta/s$. This equation 
also indicates that the modulus of quark polarization is inversely 
proportional with the center-of-mass energy. On the other hand, this 
equation is based on the one-dimensional Bjorken assumption, i.e. the 
transverse expansion was not considered, while in our model the 
spherical expansion is manifested in the second term, and obviously 
influences the final polarization significantly.

Previous experimental results, e.g. Au+Au collisions at 62.4 GeV and 
200 GeV in RHIC, have shown global polarization \cite{BIA07}, with large
error bars. We have to point out that these experiments had a centrality
percentage of 0-80\% in RHIC, which dilutes the obtained polarization 
values after averaging. Also the azimuth averaged values are
much smaller than values for given azimuthal ranges as shown in Fig.
\ref{F6ab-0y0}. Furthermore Fig. 2 in Ref. \cite{GCD08} 
and Fig. 3 in Ref. \cite{PRC08} have shown a centrality region of 
nontrivial initial angular momentum, which drops drastically above 50\% 
and below 20\% centrality percentage. Since the polarization originates 
from initial angular momentum, it is better to measure the polarization 
effect in the 20\%-50\% centrality percentage range. The centrality 
percentage value used 
in our model is 30\%, which gives us the peak value of 
inital angular momentum.

For the correct determination of the momentum space dependence of 
$\Lambda$ polarization, we have to know the reaction plane and 
the Center of Mass (CM) of the participant system in a peripheral
heavy ion reaction. The Event by Event (EbE) determination of the 
longitudinal CM of participants could be measured by the 
forward backward asymmetry of the particles in the Zero Degree
Calorimeters (ZDCs). In colliders only single neutrons are measured
in the ZDCs, so one has to extrapolate from these to the total
spectator momenta. This method to detect the EbE CM was proposed 
in Refs. \cite{Eyyubova,CseStoe2014}.

At collider experiments, e.g. the LHC-ALICE or RHIC-STAR, this 
determination was not performed up to now, with the argument
\cite{Schukraft} that nuclear multi-fragmentation may also lead
to fluctuation of single neutron hits in ZDCs, and therefore
CM frame would have been determined inaccurately.
However, at FAIR's fixed target experiments, it is possible to 
detect all the fragments from multi-fragmentation of spectators, 
thus the  CM frame can be determined accurately.  

Since the experimental measurement of global $\Lambda$
polarization is conducted around different azimuthal angle,
it is crucial to accurately define the EbE CM frame. In symmetric 
collider experiments, the CM frame de facto fluctuates around 
the actual CM frame. The fixed target FAIR setup can get rid of this
uncertainty perfectly. The Compressed Byronic Matter (CBM) 
experiments will be able to measure the polarization effects at 
SIS-100 and SIS-300 with millions times higher intensity and event
rate, up to six order of magnitude than at the RHIC Beam Energy
Scan program.

The higher multiplicity, thus allows for the high resolution 
measurement of the momentum space dependence of the $\Lambda$
polarization, which can be decisive to determine the dominant
polarization mechanism.

\section{Summary}

We have explored $\Lambda$ polarization as an observable signal for
the vorticity created in peripheral heavy ion collisions. The studies
were performed within a (3+1D) hydrodynamic simulation for U+U collisions
at FAIR energies ($\sim \sqrt{s_{NN}} = 8 $ GeV).
We predicted a sizeable polarization signature in the emitted $\Lambda$
hyperons that can directly signal the initial vorticity.
The predictions can be explored at the NICA and FAIR facilities
in the near future.

\begin{acknowledgments}

Enlightening discussions with Francesco Becattini 
are gratefully acknowledged. This work was partially supported by the 
Helmholtz International Center for FAIR within the Hessian LOEWE initiative.

\end{acknowledgments}



\begin{thebibliography}{99}

\bibitem{CseStoe2014}
L. P. Csernai and H. St\"ocker, 
J. Phys. G {\bf 41},  124001 (2014).

\bibitem{Floe2013}
S. Floerchinger and U.A. Wiedemann, 
Phys. Rev. C {\bf 88}, 044906 (2013).

\bibitem{hydro1}
  L. P. Csernai, V. K. Magas, H. St\"ocker, and D. D. Strottman,
  Phys. Rev. C {\bf 84},  024914 (2011).

\bibitem{hydro2}
  L. P. Csernai, D. D. Strottman and Cs. Anderlik,
  Phys. Rev. C {\bf 85}, 054901 (2012).

\bibitem{WCBS14}
L. P. Csernai, D. J. Wang, M. Bleicher, and H. St\"ocker,
Phys. Rev. C {\bf 90}, 021904(R) (2014).

\bibitem{CK85}
L. P. Csernai and J. I. Kapusta,
Phys. Rev. D {\bf 31} 2795 (1985).
	
\bibitem{M2001-2}
  V. K. Magas, L. P. Csernai, and D. D.
  Strottman, Phys. Rev. C {\bf 64}, 014901 (2001);
   and
  V. K. Magas, L. P. Csernai, and D. D. 
  Strottman, Nucl. Phys. A {\bf 712}, 167 (2002).

\bibitem{BNK84}
T. S. Bir\'o, H. B. Nielsen, and J. Knoll,
Nucl. Phys. B {\bf 245}, 449 (1984). 

\bibitem{Son}
  P. K. Kovtun, D. T. Son and A. O. Starinets,
  Phys. Rev. Lett. {\bf 94}, 111601 (2005).

\bibitem{CKM}
  L. P. Csernai, J. I. Kapusta, L.D. ~{McLerran},
  Phys.~Rev.~Lett.~{\bf 97},  152303 (2006).

\bibitem{CMW13}
L. P. Csernai, V. K. Magas, and  D. J. Wang,
Phys. Rev. C {\bf 87}, 034906 (2013).

\bibitem{HHW11}
X.-G. Huang, P. Huovinen, and X.-N. Wang,
Phys. Rev. C {\bf 84}, 054910 (2011).

\bibitem{PLB05}
Z.-T. Liang and X.-N. Wang,
Phys. Lett. B {\bf 629}, 20-26 (2005).

\bibitem{LPS94}
L. P. Csernai, Introduction to Relativistic Heavy Ion Collisions, 
(Jonh Wiley \& Sons Ltd., Chichester, 1994).

\bibitem{JLL16}
Y. Jiang, Z.-W. Lin, and J.-F.  Liao,
arXiv:1602.06580v1.

\bibitem{Ste2014}
J. Steinheimer, J. Auvinen, H. Petersen, M. Bleicher, H. St\"ocker, 
Phys. Rev. C {\bf 89} 054913 (2014).


\bibitem{DHBT}
L. P. Csernai, S. Velle and D. J. Wang, Phys. Rev. C {\bf 89},  034916 (2014);
L. P. Csernai, and S. Velle, Int. J. Mod. Phys. E {\bf 23}, 1450043 (2014).

\bibitem{Bec13}
F. Becattini, L. P. Csernai, and D. J. Wang,
Phys. Rev. C {\bf 88} 034905 (2013).

\bibitem{BCDG13}
F. Becattini, V. Chandra, L. Del Zanna, and E. Grossi,
Annals of Phys. {\bf 338}, 32 (2013).

\bibitem{Erratum}
F. Becattini, L. P. Csernai, D. J. Wang, and Y. L. Xie,
Phys. Rev. C {\bf 93}, 069901(E) (2016).

\bibitem{Stefan}
S. Floerchinger and U. A. Wiedemann, Journal of High Energy Physics, 
{\bf 11}, 100 (2011);  and
J. Phys. G: Nucl. Part. Phys. {\bf 38}, 124171 (2011).

\bibitem{Lisa2016}
M. A. Lisa {\it et al.} (STAR Collaboration),
Invited talk presented at the
QCD Chirality Workshop 2016, Feb. 23-26, 2016, Los Angeles, USA.

\bibitem{Db13}
A. Dbeyssi, Time-like proton form factors and heavy lepton
production at the PANDA experiment,
Proceedings of Science, (Bormio 2013) 059.

\bibitem{XGCs2015}
Y. L. Xie, R. C. Glastad, and L. P. Csernai,
Phys. Rev. C {\bf 92}, 064901 (2015).

\bibitem{BIR15}
F. Becattini, G. Inghirami, V. Rolando, A. Beraudo, 
L. Del Zanna, A. De Pace, M. Nardi, G. Pagliara, and V. Chandra,
Eur. Phys. J. C {\bf 75}, 406 (2015).

\bibitem{AM87}
A. M. Smith {\it et al.} (R608 Collaboration),
Phys. Lett. B {\bf 185}, 209 (1987).

\bibitem{GLP12}
J.-H. Gao, Z-T. Liang, S. Pu, Q. Wang, and X.-N. Wang,
Phys. Rev. Lett. {\bf 109}, 232301 (2012).



\bibitem{GCD08}
J.-H. Gao, S.-W. Chen, W.-T. Deng, and Z.-T. Liang, Q. Wang, and X.-N. Wang, 
Phys. Rev. C {\bf 77}, 044902 (2008).



\bibitem{PRL05}
Z.-T. Liang and X.-N. Wang,
Phys. Rev. Lett. {\bf 94}, 102301 (2005).


\bibitem{AP08}
F. Becattini and F. Piccinini,
Annals of Physics {\bf 323},  2452-2473 (2008).

\bibitem{PangEA2016}
L.-G. Pang, H. Petersen, Q. Wang, X.-N. Wang,
arXiv:1605.04024.

\bibitem{BIA07}
STAR Collaboration, B. I. Abelev {\it et al.},
Phys. Rev. C {\bf 76}, 024915 (2007).


\bibitem{PRC08}
F. Becattini, F. Piccinini, and J. Rizzo, 
Phys. Rev. C {\bf 77}, 024906 (2008).

\bibitem{Eyyubova}
L. P. Csernai, G. Eyyubova, and V. K. Magas,
Phys. Rev. C {\bf 86}, 024912 (2012).

\bibitem{Schukraft} 
J. Schukraft (private communication).



\end{thebibliography}
\end{document}